\documentclass[12pt]{iopart}
\eqnobysec 
\usepackage{iopams}
\usepackage{graphicx}
\usepackage{amsfonts}
\usepackage{amsmath}
\usepackage[linktocpage=true]{hyperref} 
\usepackage{lineno}
\usepackage{todonotes}
\usepackage{units}
\usepackage[latin1]{inputenc}
\usepackage{color}
\usepackage{dcolumn}
\usepackage{bm}
\usepackage{cleveref}
\usepackage{braket}
\usepackage{tikz}
\crefname{section}{§}{§§}
\Crefname{section}{§}{§§}
\usepackage{mathtools}
\usepackage{subcaption}
\usepackage{chngcntr}
\counterwithout{equation}{section} 
\usepackage[normalem]{ulem}
\usepackage{ulem}
\usepackage[version=4]{mhchem}
\usepackage[percent]{overpic}
\usepackage{cancel}
\usepackage{tabularx}
\usepackage[most]{tcolorbox}
\usepackage{ulem}

\newcommand{\p}{\partial}
\newcommand{\be}{\begin{equation}}
\newcommand{\ee}{\end{equation}}
\newcommand{\bsub}{\begin{subequations}}
\newcommand{\esub}{\end{subequations}}
\newcommand{\bea}{\begin{eqnarray}}
\newcommand{\eea}{\end{eqnarray}}
\newcommand{\ei} {\end{itemize}}
\newcommand{\bmat} {\begin{pmatrix}}
\newcommand{\emat} {\end{pmatrix}}
\newcommand{\ceff}{c_{\mathrm{eff}}}

\newcommand{\D}{\mathrm{d}}
\newcommand{\I}{\mathrm{i}}
\newcommand{\E}{\mathrm{e}}
\newcommand{\op}[1]{\ensuremath{\boldsymbol{#1}}}
\newcommand{\lagrange}[1]{\mathcal{L}_{\text{#1}}}

\usepackage{xparse}
\NewDocumentCommand{\Int}{ O{} O{} m }{\int_{#1}^{#2}\!\D #3\;}

\makeatletter
\newcommand*{\balancecolsandclearpage}{%
  \close@column@grid
  \clearpage
  \twocolumngrid
}
\makeatother

\begin{document}

\maketitle

\title{Third sound detectors in accelerated motion}

\author{Cameron R D Bunney$^{1*}$, Vitor S Barroso$^1$, Steffen Biermann$^1$, August Geelmuyden$^1$, Cisco Gooding$^1$, Gr\'egoire Ithier$^2$, Xavier Rojas$^2$, Jorma Louko$^1$ and Silke Weinfurtner$^{1,3}$}

\address{$^1$School of Mathematical Sciences, University of Nottingham, Nottingham NG7~2RD, UK}
\address{$^2$Department of Physics, Royal Holloway University of London, Egham, Surrey, TW20~0EX, UK}
\address{$^3$Centre for the Mathematics and Theoretical Physics 
of Quantum Non-Equilibrium Systems, 
University of Nottingham,
Nottingham NG7~2RD, 
UK}
\address{$^*$Author to whom any correspondence should be addressed.}
\ead{\mailto{cameron.bunney@nottingham.ac.uk}}

\vspace{10pt}
\date{March 2023, revised June 2024}

\begin{abstract}
An accelerated observer moving through empty space sees particles appearing and disappearing, while an observer with a constant velocity does not register any particles. This phenomenon, generally known as the Unruh effect, relies on an initial vacuum state, thereby unifying the experience of all inertial observers. We propose an experiment to probe this observer-dependent detector response, using a laser beam in circular motion as a local detector of superfluid helium-4 surface modes or third sound waves. To assess experimental feasibility, we develop a theoretical framework to include a non-zero temperature initial state. We find that an acceleration-dependent signal persists, independent of the initial temperature. By introducing a signal-to-noise measure we show that observing this signal is within experimental reach.
\end{abstract}

\section{Introduction}
One of the hardest-to-detect yet crucial predictions arising at the interface of quantum physics and relativity is the Unruh effect~\cite{PhysRevD.14.870}. It predicts that observers in accelerated and inertial motion perceive empty space differently. While the latter sees nothing, the observer accelerating through the quantum vacuum will experience particles popping in and out of existence with a thermal spectrum characterised by a temperature proportional to its acceleration. This disparity is fundamental for our understanding of the interplay between quantum and relativistic aspects of the Universe.

Two obstacles stand in the way of the experimental observation of the Unruh effect. The first is a technical challenge in reaching prohibitively large accelerations for an appreciable temperature. The more fundamental and as yet unexamined issue is the crucial dependence on the precise properties only provided by the quantum vacuum state, which unify the perception of all non-accelerating observers travelling at any constant velocity. For an initial state differing from the quantum vacuum, such as a thermal state, this agreement between inertial observers is tainted by a Doppler-shifted spectrum according to their velocities~\cite{Bunney}. Related previous analyses on an ambient temperature in $3+1$ dimensions are given in~\cite{ref4,BunneyRick}

In the present paper, we tackle these two issues at hand. We devise a quantum simulator in superfluid helium where the phenomenon of observer-dependence persists, and a high ambient temperature can be seized as an experimental advantage by allowing the scanning of a wide range of acceleration temperatures. Specifically, we develop a quantitative framework for both controlling this initial thermality and distinguishing between the inertial (Doppler) and acceleration (Unruh) signatures. We further show that a laser can be used as a continuous detector to extract this acceleration-dependent signal. 

An accelerated particle detector on a circular trajectory can be constructed using a localised laser beam interacting with an ultracold atoms system~\cite{PhysRevLett.125.213603,UnruhLowTemp} or by the inclusion of an impurity atom~\cite{ImpureBEC}. In previous experimental proposals using atomic clouds, the sample temperature could be ignored. However, in general, this assumption does not hold true in other quantum simulators for field theories and gravity, where finite temperatures are significant in their underlying dynamics, which we address in the present work. We emphasise that in contrast with previous experiments inferring an Unruh signature through a functional equivalence in analogue settings~\cite{ref6,ref8}, or interpretations of acceleration radiation in high-energy systems in terms of the Unruh effect~\cite{ref9,ref10,ref11,Gregori}, our proposal realises an accelerated detector by a laser beam that probes the quantum system along an accelerated worldline.

Ultracold atoms systems as gravity simulators are accessible playing fields with controllable parameter spaces and have been successful in probing fundamental phenomena~\cite{weinfurtner11,torres2017rotational,QNM,Svancara}. We show that long-wavelength perturbations on the surface of thin film superfluid helium-$4$, commonly referred to as \textit{third sound}~\cite{atkins59}, obey an effective quantum field theory~\cite{SchutzholdUnruh02, Barroso_Bunney}. In investigating the dynamics of superfluids, including both surface excitations~\cite{mcauslan2016microphotonic,harris2016laser,childress2017cavity,singh2017detecting,sachkou2019coherent,shkarin2019quantum,Baker_2016,SchechterObservation, ScholtzThirdSound,EverittThirdSound} and acoustic phonons~\cite{Lorenzo_2014,deLorenzo2017ultra,kashkanova2017superfluid,spence2021superfluid}, optomechanical systems have been effectively utilised with remarkable precision.

\begin{figure}[t!]
    \centering
    \includegraphics[width=0.67\columnwidth]{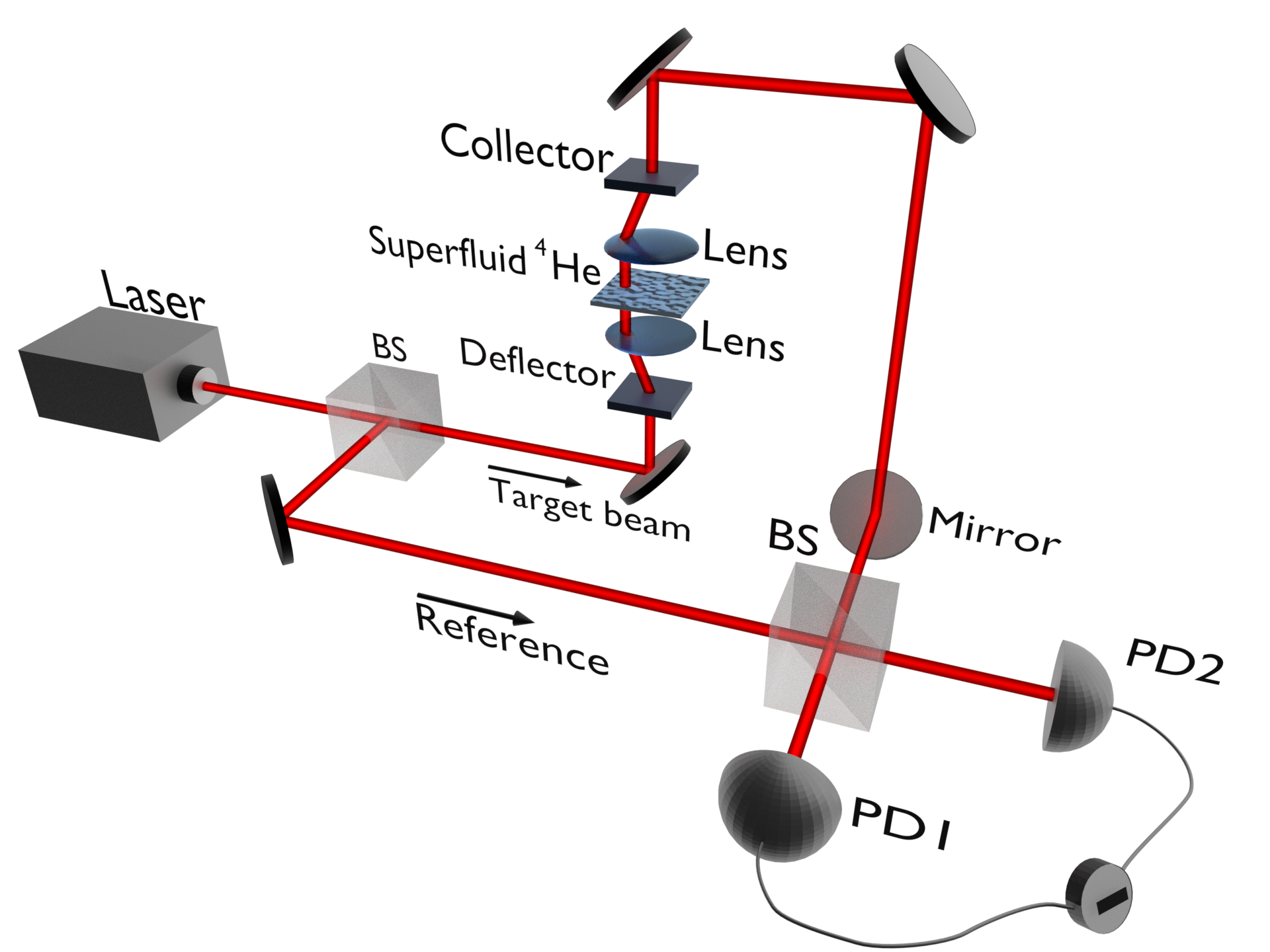}
    \caption{Experimental setup. A beam splitter (BS) separates the input laser into two arms, namely a target and a reference. The target beam is focused on the sample plane to a spot, with width much smaller than the long-wavelength surface perturbations. A deflector-lens configuration moves the target beam on a circular trajectory through the superfluid helium sample with optical axis perpendicular to the fluid surface. After the sample, a lens-collector combination, followed by a series of mirrors, brings the beam back to a static path and lead it to a second beam splitter (BS), where the target and reference arms are combined. The resulting beams are detected at two photodiodes (PD1) and (PD2).}
    \label{fig:schematics}
\end{figure}

Using a laser as a probe field, we show height fluctuations $\delta h$ on the surface of superfluid helium are transduced into phase fluctuations $\psi$ in the laser,
\begin{equation}
\psi_{\delta h} ~\approx~ \frac{1}{2}(n^2-1) k_L \delta h\,, 
\label{eq:phasetoHeight}
\end{equation}
with index of refraction $n$ and laser wavevector $k_L$. Using our proposed setup in Figure~\ref{fig:schematics}, phase-referenced (homodyne~\cite{walls08,Stefszky_2012}) photodetection retrieves the phase spectrum and with it a signature of the Unruh effect. 

In view of the central role of acceleration, we introduce the difference spectrum: a measure of the observability of acceleration-dependence in the detector response. This quantity is obtained by subtracting a purely velocity-dependent contribution arising from inertial motion through a thermal bath. We demonstrate that the difference spectrum is operationally measurable and present a signal-to-noise analysis indicating that it is observable. As such, our considerations of a realistic thermal initial state are crucial in the experimental realisation of the Unruh effect.

\section{Thin-film superfluid helium-$4$}\label{sec:thin film helium}
We begin by considering the interfacial dynamics of superfluid helium-$4$ using Landau's two-fluid model~\cite{PhysRev.60.356,DonnellyHelium}, 
which states that below a critical temperature of $2.17\,\mathrm{K}$ (the lambda-point) helium-$4$ can be decomposed into a normal component and a superfluid component. In our regime, the classical height fluctuations $\delta h$ of the superfluid component obey the $(2+1)$-dimensional Klein-Gordon equation. 
Within the temperatures we consider, and within the non-dispersive limit, canonical quantisation of these height fluctuations yields a theory of clearly-defined elementary quantum excitations (phonons), which have been well characterised \cite{mcauslan2016microphotonic,harris2016laser} and experimentally exploited \cite{sachkou2019coherent}. The experimental estimates that we give follow from this quantisation.

When deriving the equations of motion (see~\ref{app:superfluid}), we work under several assumptions. First, we will operate at low temperatures (below $500~\mathrm{mK}$), below which the superfluid component will dominate~\cite{LowTempPhys,DonnellyHelium}, and we can neglect contributions from the normal component altogether, as well evaporation and recondensation of the superfluid~\cite{atkins59}. This assumption also means that the saturated helium vapour pressure is approximately zero, and the pressure gradient originating from this interaction will vanish~\cite{ThinFilmThesis}. In addition, we will work in the isothermal limit, assuming that there is no heat transfer in the system due to wave propagation, thereby ignoring temperature gradients. Our final assumption is that we consider saturated thin films of helium-$4$ (i.e. height $h_0>10\,\mathrm{nm}$), at which thickness the superfluid component can form surface waves, which propagate independently of a stationary normal component - this is referred to as third sound~\cite{atkins59,Atkins1970Chapter2T}. Given the film thickness is orders of magnitude larger than a few atomic layers, the superfluid helium-$4$ can be considered incompressible~\cite{PhysRevB.18.2155}.

Under the aforementioned assumptions, it is shown in~\cite{Barroso_Bunney} that the dynamics of the fluid components can be effectively linearised at the superfluid surface for small perturbations in height $\delta h(t,\op{x})=h(t,\op{x})-h_0$ and velocity $\bm{v}=\bm{\nabla} \phi$, where $\phi$ is the velocity potential.
The result of this linearisation is a wave equation for height fluctuations $\delta h$,
\begin{equation}
\frac{1}{gh_0}\frac{\partial^2}{\partial t^2}\delta h~=~\nabla^2 \delta h-\frac{\sigma}{g \rho }\nabla^2(\nabla^2 \delta h) \;,
\label{eq:superfluideom}
\end{equation} where $\sigma$ is the surface tension of superfluid helium-$4$. The effective gravity $g$ contains a contribution due to gravity $g_0$ and due to van der Waals interactions of the fluid with the substrate material via the coupling constant $\alpha_{\mathrm{vdW}}$. For thin films, gravity is negligible compared to the van der Waals contribution, and the effective gravitational coupling is~\cite{Baker_2016,tilley1990}
\begin{equation}
    g~=~g_0+3\frac{\alpha_{\mathrm{vdW}}}{h_0^4}~\approx~3\frac{\alpha_{\mathrm{vdW}}}{h_0^4}\,.
\end{equation}
The dispersion relation arising from~\eqref{eq:superfluideom} is
\begin{equation}\label{eqn:dispersion}
    \omega^2~=~gh_0\left(1+\frac{\sigma}{\rho g}k^2\right)k^2\,.
\end{equation}
We consider the long-wavelength limit wherein the effective gravity dominates capillary effects, $\sigma k^2\ll g\rho$.
In this case, we linearise the dispersion relation~\eqref{eqn:dispersion} to obtain the third-sound speed (phase velocity) given by~\cite{PhysRevLett.75.3316,Baker_2016}
\begin{equation}
c_3^2 ~=~\left(\frac{\omega}{k}\right)^2~\approx~\frac{3\alpha_{\mathrm{vdW}}}{h_0^{3}}\;.
\label{eq:linearisedisothermal}
\end{equation}

\noindent In this thin-film, nondispersive limit, \eqref{eq:superfluideom} is the equation of motion for $\delta h(t,\bm{x})$ with corresponding Lagrangian
\begin{equation}
\lagrange{fluid} ~=~\frac12\rho h_0\left[\left(\frac{\partial \delta h}{\partial t}\right)^2-c_3^2|\bm{\nabla}\delta h|^2\right]\,.
\end{equation}

In this limit, we have that the height fluctuations $\delta h$ and velocity potential $\phi$ each obey a Klein-Gordon equation with propagation speed $c_3$ with a linear dispersion $\omega=c_3k$~\cite{Barroso_Bunney}. In this effective field theory, the height field is the conjugate momentum to the fundamental field, the velocity potential $\phi$, which obey the relation
\begin{equation}\label{eqn:conjugate momentum}
    \delta h(t,\bm{x})~=~-\frac{1}{g}\frac{\partial}{\partial t}\phi(t,\bm{x})\,.
\end{equation}
This is identical in form to the equations governing a quantum scalar field used in the quantum field theoretical (QFT) derivation of the Unruh effect. This correspondence laid the groundwork for the birth of analogue gravity~\cite{PhysRevLett.46.1351}, to which we appeal to motivate this work.

\section{Lasers as local detectors of interface fluctuations}\label{sec:lasers}
For a $(3+1)-$dimensional electromagnetic field, propagating in the $z$-direction with only one polarisation, we can express the electromagnetic potential as $A(t,z)=A_0\cos\left(\omega_L t-k_Lz+\psi(t,z)\right)$. Before interaction with the superfluid helium-4, the quantised field $\psi$ is electromagnetic noise. We take the background laser field to be a plane wave, $A_0 \cos (\omega_Lt - k_L z)$, and the perturbation $\psi(t,z)$ to be a real field describing the phase fluctuations of the laser.
By linearising the standard electromagnetic Lagrangian in terms of small phase perturbations and after an appropriate rescaling of $\psi$ as derived in~\cite{PhysRevLett.125.213603}, we obtain
\be 
  \lagrange{em} ~=~ \frac{1}{2 \mu_0}\left(\frac{1}{c^2}\left(\frac{\partial}{\partial t} \psi(t,z)\right)^2-\left(\frac{\partial}{\partial z} \psi(t,z)\right)^2\right)\;,
\ee where $\mu_0$ is the vacuum permeability and $c$ is the speed of light in vacuum.

When the laser beam passes through the superfluid, the atoms will react by forming dipoles according to their polarisabilities $\alpha$. Assuming the laser is sufficiently detuned from atomic resonances, $\alpha$ can be taken to be real and the superfluid-light interaction can be calculated within a semiclassical model in the framework of macroscopic electrodynamics. To shorten the notation in the following, $\alpha$ refers to the real part of the polarisability. The interaction between the superfluid and the laser is
\be
\lagrange{int} ~=~ \frac{1}{2}\alpha\rho_N\left(\partial_tA\right)^2\;,
\label{eq:intL}
\ee  
with $\rho_N$ the $3\mathrm{D}$ number density of helium-$4$~\cite{Agarwal2014}. Using the above ansatz for $A$, this Lagrangian \eqref{eq:intL} can be expanded in small $\psi$. In \ref{app:emforce}, we derive the equation of motion for the laser phase localised to the trajectory $\bm{x}=\bm{X}(t)$ at the surface, where we will consider the interaction to be localised to a region on the free surface of the superfluid, lying on a circular trajectory with radius $R>0$ and angular frequency $\Omega>0$. This path can be parametrised as $\bm{X}(t)=(R\cos(\Omega t),R\sin(\Omega t))$. The equation of motion reads
\begin{equation}\label{phase eom}
\Box\psi(t,z)~=~-\alpha\rho_N\omega_L\mu_0\partial_t\delta h(t,\bm{X}(t))\delta(z-h_0)\,,
\end{equation}
where $\Box=(-\ceff^{-2}\p_t^2+\p_z^2)$ is the wave operator and $\ceff$ is the effective speed of light, which models the effective slow-down of light through the superfluid via the electrostrictive interaction and is obtained from the interaction Lagrangian \eqref{eq:intL}. 

The equation of motion for the phase field $\psi$ in the interacting laser-superfluid system~\eqref{phase eom} is backreactive: there is a contribution due to the height fluctuations. On the other hand, the equation of motion for the height field $\delta h$ contains a contribution from the phase field $\psi$. In the limit of negligible back action of the laser phase on the height, the homogeneous part of~\eqref{phase eom} gives a solution $\psi_0(t,z)$, whereas the inhomogeneous part can be solved using the Green's function for the differential operator $\Box$, given by $G(t,z)=\frac{\ceff}{2}\Theta(t-|z|/\ceff)$, where $\Theta(x)$ is the Heaviside function. This results in the solution
\begin{equation} \label{eq:psieomsolution}
\psi(t,z)~=~\psi_0(t,z) -\frac{1}{2}\alpha\rho_N\omega_L\mu_0\ceff\delta h(\tau,\bm{X}(\tau)),
\end{equation} 
where $\tau = t-|z-h_0|/\ceff$. It is then clear that the laser phase contains a contribution from the height fluctuations $\delta h$ sampled along the interaction trajectory $\bm{X}(\tau)$. The phase fluctuation for weak interactions attributable to the height fluctuations given by (\ref{eq:psieomsolution}) can be written as
\begin{equation}
\label{eq:sol}
\psi_{\delta h} ~=~ \frac{1}{2}\alpha \rho_N \omega_L \ceff \mu_0\delta h ~=~ \frac{1}{2n}(n^2-1)k_L \delta h\, ,
\end{equation}
in terms of an index of refraction $n$ as defined by
\be 
 n~=~\frac{c}{c_{\text{eff}}} ~=~\sqrt{1+\frac{\alpha\rho_N}{\varepsilon_0}} \;,
\label{eq:indexofrefraction}
\ee where $\varepsilon_0$ is the vacuum permittivity. Note that \eqref{eq:sol} differs from the phase associated with an optical path-length difference and is approximated in \eqref{eq:phasetoHeight} for $n$ close to unity.

Due to the interaction~\eqref{eq:intL}, the laser phase samples unequal time and unequal space correlations in the height field along the interaction trajectory. As a result, the characteristic observer dependence of Unruh-like phenomena is encoded in the phase correlations of the outgoing beam. As illustrated in~\Fref{fig:schematics}, an interferometric scheme between reference and target beams retrieves the phase fluctuations from the readouts of photodiodes $1$ and $2$.

\section{Uniform circular motion response}\label{sec:circmot}
As the laser samples fluctuations of the helium surface along the interaction trajectory, the laser acts as an effective continuous field Unruh-DeWitt (UDW) detector~\cite{PhysRevLett.125.213603}. The response of a UDW detector is the Fourier transform of the two-point correlation function, the power spectral density (PSD). For the phase fluctuations $\psi_{\delta h}$, the PSD at frequency $\omega$ is given by
\begin{equation}
    S(\omega)~=~\int_{-\infty}^\infty\D\tau\,\E^{-\mathrm{i}\omega\tau}\braket{\psi_{\delta h}(\tau)\psi_{\delta h}(0)}_T~=~\frac{(n^2-1)^2k_L^2}{4}\int_{-\infty}^\infty\D\tau\,\E^{-\mathrm{i}\omega\tau}\braket{\delta h(\tau)\delta h(0)}_T\,,
\end{equation}where $\braket{\psi_{\delta h}(\tau)\psi_{\delta h}(0)}_T$ is the thermal correlation function along the trajectory $\bm{X}$ at temperature $T$ and we have used~\eqref{eq:phasetoHeight}.

 As the height field $\delta h$ is proportional to the time derivative of the velocity potential $\phi$ of the superfluid~\eqref{eqn:conjugate momentum}, we rewrite this PSD as
\begin{equation}
    S(\omega)~=~\frac{(n^2-1)^2k_L^2}{4g^2}\int_{-\infty}^\infty\D\tau\, \E^{-\mathrm{i}\omega\tau}\braket{\left.\tfrac{\D}{\D \tau'}\phi (\tau')\right|_{\tau'=\tau}\left.\tfrac{\D}{\D \tau''}\phi (\tau'')\right|_{\tau''=0}}_T\,.
\end{equation}The integrand is the twice-differentiated thermal Wightman function which eliminates the infrared divergence that would occur in the undifferentiated thermal Wightman function~\cite{Bunney}.

Using~\eqref{eq:psieomsolution}, the PSD for the full phase field $\psi$ can be decomposed as
\begin{equation}\label{psdmeas}
    S_{\psi}(\omega) ~=~ \kappa\mathcal{F} (\omega;T)+\sigma_{\mathrm{sn}}^2\,, 
\end{equation} 
where $\sigma_{\mathrm{sn}}^2$ comes from the shot noise in the signal~\cite{PhysRevLett.125.213603}, while $\mathcal{F}(\omega;T)$, known as the response function in the literature, comes from the analogue field fluctuations. The dimensionful constant $\kappa$ in \eqref{psdmeas} arises from matching the normalisation of our fields to that in~\cite{Bunney} (see~\ref{app:eft}), and is given by 
\begin{align}\label{kappadef}
    \kappa ~=~ \frac{\hbar^2(n^2-1)^2k_L^2}{4\rho g}~\approx~ 0.51\times10^{-64}\,\mathrm{kg}\,\mathrm{m}^{4}\,,
\end{align} 
where the numerical value is calculated from the parameters given in Table~\ref{tab:params}.

Specialising to uniform circular motion with constant velocity $v$ and acceleration $a$, the response function is given by~\cite{Bunney}
\begin{multline}\label{eqn::circmotion}
    \mathcal{F}(\omega;T)~=~\frac{\omega^2}{2\hbar c_3^2}\sum_{m>\omega v/a}J^2_{m}\left(\frac{mv}{c_3}-\frac{\omega v^2}{ac_3}\right)
    \\+\frac{\omega^2}{2\hbar c_3^2} \sum_{m>\omega v/a}\frac{1}{\E^{\hbar(ma/v-\omega)/k_BT}-1}J^2_{m}\left(\frac{mv}{c_3}-\frac{\omega v^2}{ac_3}\right) \\+\frac{\omega^2}{2\hbar c_3^2}\sum_{m>-\omega v/a}\frac{1}{\E^{\hbar(ma/v+\omega)/k_BT}-1}J^2_{m}\left(\frac{mv}{c_3}+\frac{\omega v^2}{ac_3}\right)\,, 
\end{multline}
where $J_m$ are the Bessel functions of the first kind~\cite{NIST}. 
Eq.~\eqref{eqn::circmotion} is the thermal response function for a massless scalar field with an interaction that includes a time derivative in the coupling, as discussed in~\cite{Bunney}.

\subsection{Linear motion}

The circular motion response function 
$\mathcal{F}(\omega;T)$ \eqref{eqn::circmotion} encodes two distinct motion effects: one is due to the detector's acceleration, while the other is a Doppler effect, due to the detector's speed $v$ with respect to the ambient heat bath. 
To identify the parameter regime in which the acceleration effect is significant compared with the Doppler effect, we compare $\mathcal{F}(\omega;T)$ 
with the response function of a detector in non-accelerated, linear motion with the same speed~$v$, wherein only the Doppler effect is present. 
The linear motion response function is~\cite{Bunney}
\begin{align}\label{deffl}
   \mathcal{F}_{\mathrm{Lin}}(\omega;T)=\frac{1}{2}\frac{\omega^2}{\hbar c_3^2}\gamma_s\Theta\left(-\frac{\omega}{c_3}\right)
   +\frac{|\omega|^2\gamma_s}{2\pi\hbar c_3^2} \int_{-\pi/2}^{\pi/2}
    \frac{\mathrm{d}\theta}{\E^{\gamma_s^2(1 + \frac{v}{c_3}\sin\theta)\hbar|\omega|/k_BT}-1}\,,
\end{align}
where $\gamma_s=(1-v^2/c_3^2)^{-1/2}$ is the Lorentz factor. 
For $\gamma_s^2 \hbar |\omega|/k_B T \ll 1$, which holds in our parameter regime, 
\eqref{deffl} reduces to 
\begin{align}\label{deffl-smallomega}
   &\mathcal{F}_{\mathrm{Lin}}(\omega;T)=\frac{1}{2}\frac{\omega^2}{\hbar c_3^2}\gamma_s\Theta\left(-\frac{\omega}{c_3}\right) 
   + \frac{|\omega|k_B T}{2\hbar^2 c_3^2} \,, 
\end{align}
where the second term is linear in $|\omega|$ and independent of~$v$.

\section{Experimental estimates}
We have seen in~\Sref{sec:lasers} that fluctuations in the height field induce fluctuations in the laser phase. In this Section, we identify the acceleration dependence in the total PSD $S_\psi$~\eqref{psdmeas} due to the circular motion of the laser beam. The acceleration effect has been shown to be approximately thermal~\cite{bell83,bell87,UNRUH1998163} and is characterised over most of the parameter space by an approximate Unruh temperature $T_U$, given by~\cite{PhysRevLett.125.213603,PhysRevD.102.085006}
\be 
\label{UnruhTemp}
k_B T_U ~=~ \frac{\gamma_s v^2}{2\pi c_3^2} \hbar \Omega_R\;,
\ee 
where $c_3$ is the third sound speed \eqref{eq:linearisedisothermal}, 
$v$ is the orbital speed of the laser dot on the helium surface, 
$\Omega_R=c_3/R$ and $\gamma_s=(1-v^2/c_3^2)^{-1/2}$. 
We show that the circular acceleration effect is identifiable 
in the PSD, even when $T_U$ is smaller than the helium sample temperature. 
We emphasise that our system is not subject to the 
difficulties in isolating the circular acceleration effect in synchrotron electron beam depolarisation measurements, as in~\cite{bell83,bell87,UNRUH1998163}. 
\newcolumntype{b}{X}
\newcolumntype{s}{>{\hsize=.5\hsize}X}
\begin{table}[t!]
\caption{Experimental parameters for helium-4. The van der Waals constant is given for a quartz substrate~\cite{PhysRevLett.75.3316}; for other materials, $\alpha_{\mathrm{vdW}}$ would be of the same order of magnitude.}
\label{tab:params}
\begin{center}
\begin{tabularx}{\the\columnwidth}{ 
   >{\raggedright\arraybackslash}b
   >{\centering\arraybackslash}b
   }
\hline
\rule{0pt}{10pt}\textbf{Parameter} & \textbf{Value}\\ \hline
Film height, $h_0$ & $100~\textrm{nm}$ \\
Mass density, $\rho$ & $145~\textrm{kg}\,\textrm{m}^{-3}$ \\
Surface tension, $\sigma$ & $37.9\times 10^{-5}~\textrm{J}\,\textrm{m}^{-2}$ \\
Van der Waals const., $\alpha_{\mathrm{vdW}}$ & $2.6\times10^{-24}~\mathrm{m}^5\mathrm{s}^{-2}$ \\
Index of refraction, $n$ & 1.025\\
\hline 
\end{tabularx}
\end{center}
\end{table}
Having established the prerequisite formalism, we move to discuss experimentally realisable scenarios. In our gravity simulator, we are able to tune the third sound speed $c_3$ of our helium sample by varying the depth $h_0$ (cf.~\eqref{eq:linearisedisothermal}). We consider a typical saturated film thickness of $h_0=100\,\textrm{nm}$, which is within experimental reach (experimental parameters are listed in Table~\ref{tab:params})~\cite{PhysRevLett.75.3316,harris2016laser,hoffmann2004measurements} and corresponds to a third sound speed $c_3 \approx 8.8\times 10^{-2}\; \textrm{m}\, \textrm{s}^{-1}$. We recall that we operate in the nondispersive regime $\omega\ll\tfrac{1}{2} c_3\sqrt{\rho g/\sigma}$, which for the current parameters corresponds to $\tfrac{1}{2} c_3\sqrt{\rho g/\sigma}\approx (2\pi )1.2\,\mathrm{kHz}$. 

In order to achieve an analogue-relativistic speed of $v=0.95c_3$ with rotational frequencies of the order of hundreds of Hertz, we specify a trajectory radius $R=60\,\mu\textrm{m}$. These parameters correspond to an Unruh temperature~\eqref{UnruhTemp} $T_U\approx 5\,\textrm{nK}$. Standard superfluid helium-$4$ operating temperatures are orders of magnitude above $T_U$, in the range of milli-Kelvin to Kelvin temperatures~\cite{DonnellyHelium,harris2016laser}. As such, the laser, even when static, will pick up thermal fluctuations of the sample.

In line with our assumptions, we consider a helium-sample temperature of $T=10\,\mathrm{mK}$. Modelling the superfluid surface by a relativistic field prepared in a thermal state as in~\Sref{sec:circmot}, the PSD of the height fluctuations sampled by the laser is related to the UDW detector response function $\mathcal{F}$~\eqref{eqn::circmotion} through~\eqref{psdmeas}~\cite{Bunney}. The response function $\mathcal{F}$~\eqref{eqn::circmotion} splits into a contribution from the vacuum and a contribution from the ambient temperature. As $T\gg T_U$, the response function is dominated by the finite temperature contribution. We note that purely classical thermal fluctuations of the surface waves are discussed in~\cite{Flekkoy95,Flekkoy96}.

\begin{figure}[t!]
    \centering
    \includegraphics[width=0.5\columnwidth]{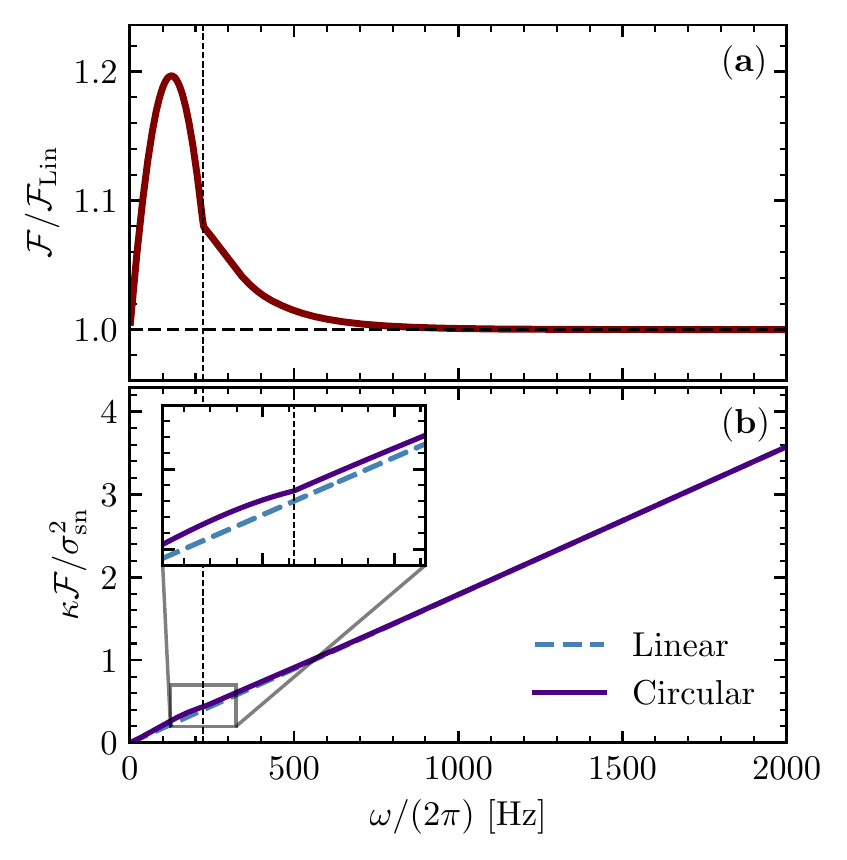}
    \caption{(a) Ratio of circular motion response function \eqref{eqn::circmotion} to linear motion response function~\eqref{deffl-smallomega}. (b) Ratio of response to shot noise, for both circular motion and linear motion. 
    The plots are at temperature $T=10\,\mathrm{mK}$. }
    \label{fig: flin comparison}
\end{figure}

In order to isolate the acceleration dependence in the detector response, we define the difference spectrum $S_\delta(\omega)$~\cite{Bunney},
\begin{equation} \label{defNDS}
S_\delta(\omega) ~=~ S_{\psi}(\omega)-\sigma_{\mathrm{sn}}^2-\kappa\mathcal{F}_{\mathrm{Lin}}(\omega;T)\, ,
\end{equation}
where shot-noise-induced phase fluctuations are approximated as Gaussian with variance $\sigma^2_{\mathrm{sn}}=\hbar\omega_L/P$, assumed independent of frequency within the measurement band. The difference spectrum $S_\delta$~\eqref{defNDS} is a measure of the deviation between the detector response on an accelerated trajectory $\mathcal{F}$ and on a linear trajectory $\mathcal{F}_{\mathrm{Lin}}$, both at constant speed $v$. As such, $S_\delta$ extracts the acceleration dependence in the response. We note that by direct calculation, the difference spectrum for a static detector ($v=0$) vanishes, $S_\delta(\omega)=0$.

\Fref{fig: flin comparison} compares the circular motion and linear motion response functions over our frequency range. As the two closely agree at high frequencies,  
a measurement of $\mathcal{F}(\omega;T)$ at high frequencies determines $\mathcal{F}_{\mathrm{Lin}}(\omega;T)$ over our full frequency range, by the linearity of~\eqref{deffl-smallomega}. As such the quantity $\mathcal{F}_{\mathrm{Lin}}$ in~\eqref{defNDS} is operationally measurable from the circular-motion response \eqref{psdmeas} alone.

\begin{figure*}[t!]
\begin{subfigure}{0.5\columnwidth}
  \centering
  \includegraphics[height=8.5cm]{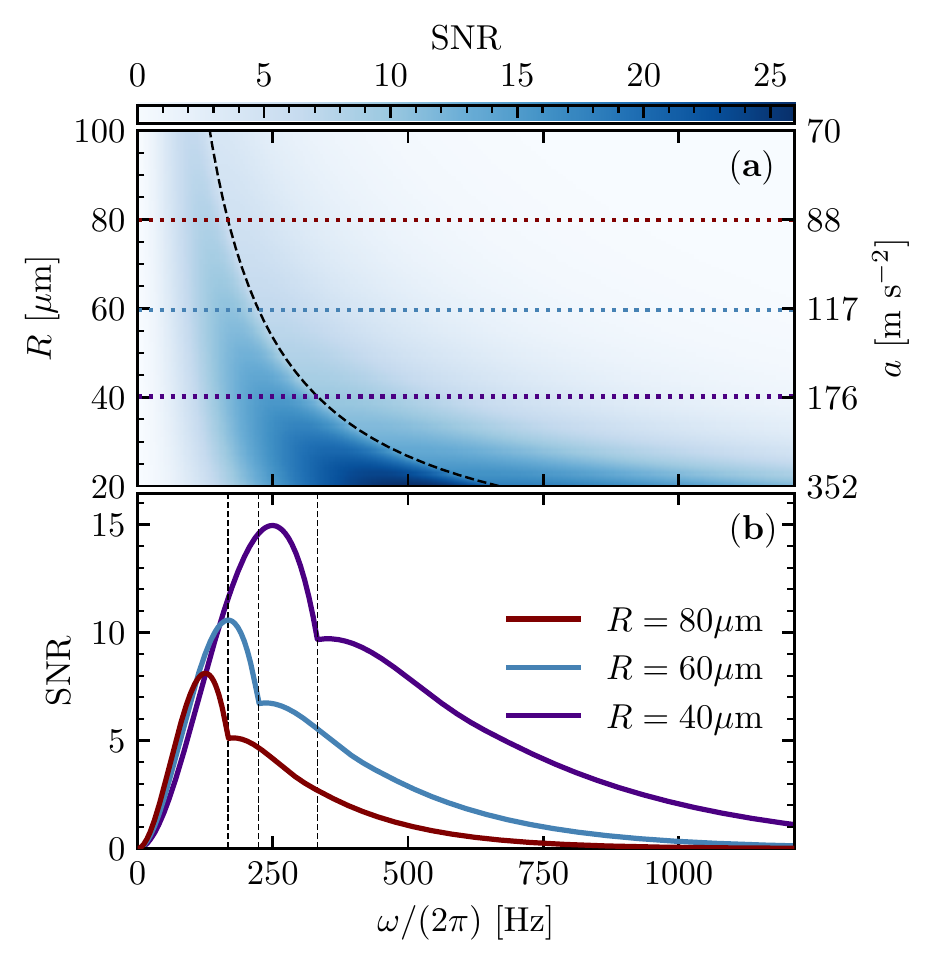}
  \caption[i]{}
    \label{fig:NDS}
\end{subfigure}
\begin{subfigure}{0.5\columnwidth}
  \centering
  \includegraphics[height=8.5cm]{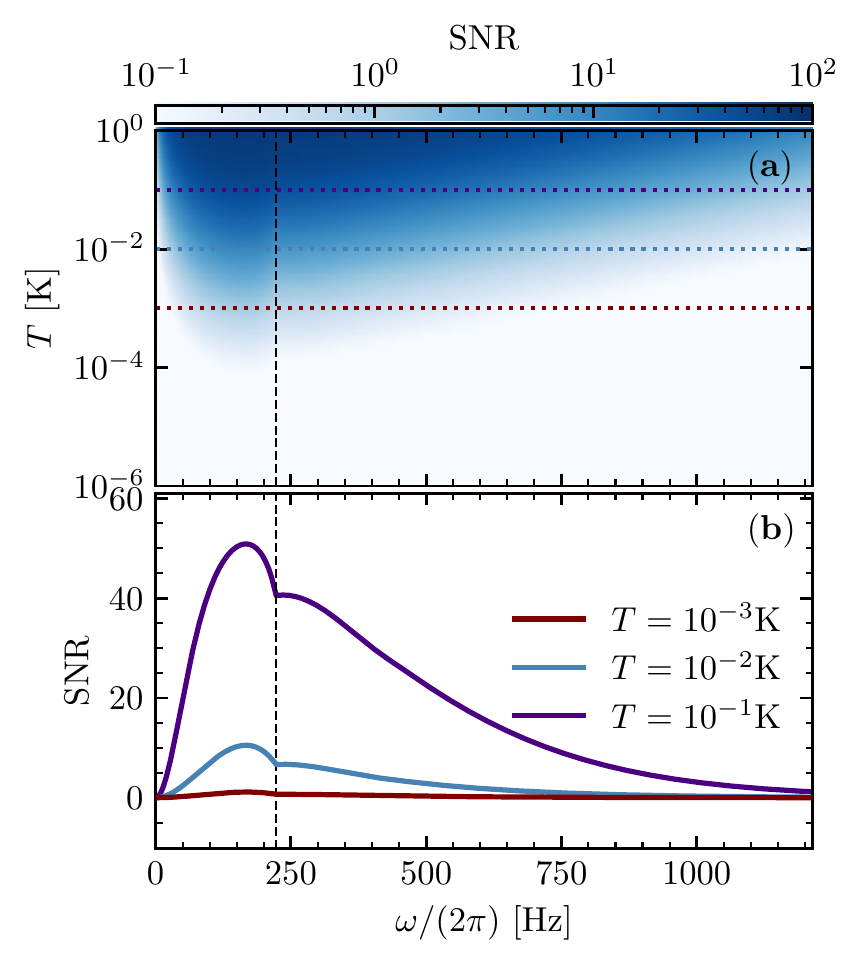}
    \caption[ii]{}
    \label{fig: NDSTemp}
\end{subfigure}
\caption{\textbf{(i)} SNR \eqref{SNR} for $\mathcal{N}=10^5$ realisations with a bandwidth of $\mathcal{B}=1$. Superfluid temperature $T=10\,\mathrm{mK}$, orbital speed $v=0.95c_3$, orbital radius $R$ as shown; the laser trajectory angular frequency and acceleration are determined in terms of these quantities by $\Omega = v/R$ and $a = v^2/R$. The horizontal axis is the frequency. Panel (a) Heatmap of computed SNR for orbital radii $20\,\mu\mathrm{m}\leq R\leq100\,\mu\mathrm{m}$. The horizontal colorbar indicates the magnitude of the SNR. Dashed black curve is the orbital angular frequency $\Omega$. Purple, blue and dark red dotted lines are lines of constant radii, whose SNRs are displayed in panel (b). Panel (b) Profile of computed SNR for constant radial slices in panel (a). Vertical black dashed lines represent orbital angular frequencies for $R=40\,\mu\mathrm{m}$ (far right), $60\,\mu\mathrm{m}$, and $80\,\mu\mathrm{m}$ (far left). \textbf{(ii)} SNR \eqref{SNR} for $\mathcal{N}=10^5$ realisations with a bandwidth of $\mathcal{B}=1$. Orbital radius $R=60\,\mu\mathrm{m}$, speed $v=0.95 c_3$, frequency on the horizontal axis as in Figure~\ref{fig:NDS}. Dashed vertical black line shows the orbital angular frequency $\Omega$. Panel (a) Heatmap of computed SNR for superfluid at temperatures $1\,\mu\mathrm{K}\leq T\leq 1\,\mathrm{K}$. Purple, blue and dark red dotted lines are lines of constant superfluid temperature, whose SNRs are displayed in panel (b). Panel (b) Profile of computed SNR for constant temperature slices in panel (a).}
\end{figure*}

To determine the feasibility of measuring the difference spectrum, we use the signal-to-noise (SNR) quantifier
\begin{equation}\label{SNR}
\mathrm{SNR}~=~ \sqrt{\frac{\mathcal{N}\mathcal{B}}{2}}\frac{S_\delta(\omega)}{\sigma^2_{\mathrm{sn}}\sqrt{1+2\frac{S_\delta(\omega)}{\sigma^2_{\mathrm{sn}}}+2\frac{S_\delta^2(\omega)}{\sigma^4_{\mathrm{sn}}}}}\, ,
\end{equation}
where $\mathcal{N}$ is the number of realisations and $\mathcal{B}$ is the resolution bandwidth in units of the measurement bandwidth~\cite{PhysRevLett.125.213603}. In our calculations, we assume a laser power $P=0.5\,\mathrm{mW}$, which is low enough to prevent the evaporation of the superfluid helium-$4$, with wavelength $\lambda_L=700\,\mathrm{nm}$ and beam width $r_0=10\,\mu\mathrm{m}$. Figures \ref{fig:NDS} and \ref{fig: NDSTemp} explore the wider parameter space for viable experimental implementation. For a constant velocity $v$, the acceleration $a=v^2R^{-1}$ is inversely proportional to the orbital radius $R$; hence, Figure \ref{fig:NDS} can be interpreted in terms of the Unruh effect: the signal increases with acceleration. The blue curve in Figures \ref{fig:NDS} (b) and \ref{fig: NDSTemp} (b) represent the common intersection of the two parameter searches. Figure \ref{fig: NDSTemp} shows that the acceleration-dependent features are amplified by an increase in the temperature of superfluid helium. This is also noted in our supporting paper~\cite{Bunney}, wherein this increase in signal with increasing ambient temperature is interpreted as a lowering in the effective temperature experienced by the detector.

\section{Conclusion}
We have proposed an experimental setup that exhibits an analogue of the circular-motion Unruh effect in thin film superfluid helium-4. Using a continuous probing field (i.e. a laser) as a detector, one can sample surface fluctuations along an accelerated (circular) interaction trajectory. In the third sound regime, the surface fluctuations behave as an effective relativistic field, and the detector response function extracted from the laser phase carries information about the acceleration along the circular path. We note that the analysis presented here does not account for either finite-size effects or the induced noise in the superfluid due to the electrostrictive drag of the laser. We leave an in-depth examination of these effects to future work. 

Using a signal-to-noise measure derived from the principle of extracting only acceleration-dependent effects, we explored experimentally viable superfluid helium-4 temperatures and rotating detector radii. 
The results in Figures \ref{fig:NDS} and \ref{fig: NDSTemp} correspond to a film thickness of $100\,\mathrm{nm}$ and show the behaviour of the system in the high-temperature limit, $k_B T\gg \hbar \omega$, for frequencies $\omega$ within the nondispersive band. By reducing the film thickness, one can shift the SNR peak - and broaden the nondispersive band - to high enough frequencies for vacuum effects to potentially be visible (i.e. $k_B T \lesssim \hbar \omega$).

We re-emphasise that the initial state in our helium-$4$ system is thermal, and significantly different from the vacuum that features in textbook descriptions of the Unruh effect, but we found that an acceleration-dependent response persists, with the SNR increasing with acceleration. Furthermore, we showed that the Unruh signature is amplified as the temperature of the superfluid helium-$4$ increases. In this way, our proposed experimental setup, framework, and analysis are pivotal and provide a stepwise approach, enabling the immediate experimental exploration of the circular-motion Unruh effect.

\ack

We thank J\"org Schmiedmayer, Bill Unruh, Sebastian Erne, and other members of the Quantum Sensors discussion group, for helpful discussions. 
We thank Radivoje Prizia for producing the $3\mathrm{D}$ schematic in Figure \ref{fig:schematics}.
We thank anonymous referees for helpful comments. 
We acknowledge support provided by the Leverhulme Research Leadership Award (RL-2019- 020),
the Royal Society University Research Fellowship
(UF120112, UF150140, URF\textbackslash R\textbackslash 211009, RF\textbackslash ERE\textbackslash 210198) and the Royal Society Enhancements Awards and Grants
(RGF\textbackslash EA\textbackslash 180286, RGF\textbackslash EA\textbackslash 181015, RGF\textbackslash EA\textbackslash 180099, RGF\textbackslash R1\textbackslash 180059, RPG\textbackslash 2016\textbackslash 186), and partial
support by the Science and Technology Facilities Council (Theory Consolidated Grant ST/P000703/1), the Science and Technology Facilities Council on Quantum Simulators for Fundamental Physics (ST/T006900/1, ST/T005998/1) as part
of the UKRI Quantum Technologies for Fundamental
Physics programme.
The work of JL was supported by United Kingdom Research and Innovation Science and Technology Facilities Council [grant number ST/S002227/1]. For the purpose of open access, the authors have applied a CC BY public copyright licence to any Author Accepted Manuscript version arising.

\appendix

\section{Linearisation of superfluid interfacial dynamics\label{app:superfluid}
}
In this Appendix, we derive the linearised equations of motion for the helium free surface. We consider a film of superfluid helium confined in a container with a flat bottom at $z=0$ and a free surface at $z=h(t,\bm{x})$. We assume a constant, uniform density $\rho$ and align the vertical axis such that $\bm{g}=-\bm{\nabla} (gz)$. Imposing the no-penetration boundary condition at $z=0$ and the kinematic boundary condition at the free surface allows the linearisation of the fluid dynamical equations in terms of small height perturbations $\delta h(t,\op{x})=h(t,\op{x})-h_0$~\cite{Atkins1970Chapter2T}. The governing equations can be further simplified by the assumptions stated in~\Sref{sec:thin film helium}. The velocity field of the superfluid component $\op{v}_s$ is assumed to be irrotational such that $\bm{\nabla}\times \op{v}_s=0$, and therefore $\op{v}_s=\bm{\nabla}\phi$.

We impose the kinematic boundary condition at the helium-vapour-helium interface at $z=h(t,\bm{x})$ and the no-penetration boundary condition at $z=0$ for fluid components $v_i$,
\begin{align}
    v_z|_{z=0}~=~&0\,,\\
    v_z|_{z=h}~=~&\frac{\partial h}{\partial t}+\bm{v}|_{z=h}\cdot\bm{\nabla} h\,.
\end{align}
We utilise these boundary conditions to linearise the governing equations in terms of small height perturbations $\delta h(t,\op{x})=h(t,\op{x})-h_0$~\cite{Atkins1970Chapter2T},
\begin{align}
    \rho \frac{\partial \delta h}{\partial t} +\rho h_0\op{\nabla}\cdot \op{v}_s + J_m^{\mathrm{vap}} ~=~ & 0 \label{eq:superfluid1}\,,\\
    \frac{\partial \op{v}_s}{\partial t} + g\op{\nabla} \delta h -\frac{\sigma}{\rho}\op{\nabla}(\nabla^2 \delta h )+ \frac{1}{\rho}\op{\nabla} p - s \op{\nabla} T ~=~ & 0\,,  \label{eq:superfluid2}
\end{align}
where $J_m^{\mathrm{vap}}=\D m/\D t$ the mass variation due to evaporation and recondensation, $\sigma$ the surface tension, $s$ is the entropy of the fluid, and $g$ is the effective gravity. For thin films, the effective gravitational coupling is dominated by its contribution from the van der Waals interaction of the fluid with the substrate material~\cite{Baker_2016,tilley1990},
\be
g ~=~ g_0 +3\frac{\alpha_{\mathrm{vdW}}}{h^4} ~\approx~ 3\frac{\alpha_{\mathrm{vdW}}}{h^4}\,,
\ee
where $g_0$ is the usual acceleration due to gravity and $\alpha_{\mathrm{vdW}}$ is the van der Waals coupling (see~Table~\ref{tab:params}).

Working within the isothermal limit, we can neglect the temperature gradient $\bm{\nabla} T=0$~\cite{atkins59}. By considering low temperatures, we neglect the pressure gradient $\bm{\nabla} p=0$, and the evaporation/recondensation on the surface $J_m^{\mathrm{vap}}=0$~\cite{atkins59}. We then simplify equations~\eqref{eq:superfluid1} and \eqref{eq:superfluid2} and combine them into a wave equation for height fluctuations $\delta h$,
\be 
\frac{1}{gh_0}\frac{\partial^2}{\partial t^2}\delta h~=~\nabla^2 \delta h-\frac{\sigma}{g \rho }\nabla^2(\nabla^2 \delta h) \;.
\ee 

\noindent In this limit, the speed of surface wave propagation is given by~\cite{PhysRevLett.75.3316}
\be
c_3^2 ~=~ \frac{\omega^2}{k^2} ~=~ \left(gh_0+ \frac{1}{\rho}\sigma h_0 k^2\right)\;.
\label{eq:energyisothermal}
\ee
For thin films, where the effective gravity dominates over capillary effects, i.e., $\sigma k^2\ll g\rho$ in \eqref{eq:energyisothermal}, the dispersion is approximately linear $\omega\approx\sqrt{gh_0}k$, with third sound speed
\be 
c_3 ~=~ \sqrt{\frac{3\alpha_{\mathrm{vdW}}}{h_0^3}}\;.
\ee 

\section{Electromagnetic interaction and equations of motion\label{app:emforce}}
In this Appendix, we present the light-matter interaction between the laser probe and the superfluid helium sample. By a variational principle, we find the equations of motion governing this interacting system and their solution.

The linearised electromagnetic Lagrangian is given by~\cite{PhysRevLett.125.213603}
\be 
  \lagrange{em} ~=~ \frac{1}{2 \mu_0}\left(\frac{1}{c^2}\left(\partial_t \psi(t,z)\right)^2-\left(\partial_z \psi(t,z)\right)^2\right)\;.
\ee The light-matter interaction with dielectric permittivity $\varepsilon ~\approx~ \varepsilon_0+\rho_N\alpha$ is given by the potential
\be 
V ~=~ -\frac{\alpha}{2}\rho_N(\partial_tA)^2 \;,
\label{eqn:empotential2}
\ee 
where $\rho_N$ and $\alpha$ are the number density and the polarisability of helium-$4$ respectively.

We take the ansatz $A(t,z)=A_0\cos(\omega_Lt-k_Lz+\psi(t,z))$, average over high frequencies and absorb a factor of $A_0/\sqrt{2}$ into $\psi$, we then express the electromagnetic interaction as
\begin{equation}
    \lagrange{int} ~=~ \frac{\alpha\rho_N}{2} \left(\frac{A_0\omega_L}{\sqrt{2}}+\partial_t\psi\right)^2\;.
    \label{eqn:linempotential}
\end{equation}
This can be split into three parts corresponding to powers in $\psi$. We momentarily drop the constant term. Then, the terms linear and quadratic in $\psi$ are given by
\begin{equation}\label{deltaV}
     \lagrange{int}^{(1)} +\lagrange{int}^{(2)} ~=~\alpha\rho_N \frac{A_0\omega_L}{\sqrt{2}}\partial_t\psi +\frac{\alpha\rho_N}{2}\left(\partial_t\psi\right)^2\;.
\end{equation}
The quadratic term contributes to the propagation speed of $\psi$, which can be seen by combining it with the time derivative in the free Lagrangian to obtain 
\begin{align}
\lagrange{em} &~=~ \left(\frac{1}{2\mu_0}\frac{1}{c^2}+\frac{\alpha\rho_N}{2}\right) \left(\partial_t \psi(t,z)\right)^2-\frac{1}{2\mu_0}(\p_z\psi(t,z))^2\,, \\
&~=~ \frac{1}{2\mu_0}
\left(
\frac{1}{c^2}
\left(1+\alpha\rho_N c^2\mu_0\right)
\left(\partial_t \psi(t,z)\right)^2-(\p_z\psi(t,z))^2 \right)\;,\label{fullem}
\end{align}
which defines an effective speed of light in the medium 
\be \label{effectiveC}
\frac{1}{c_{\text{eff}}^2} ~=~ \frac{1}{c^2}\left(1+\frac{\alpha\rho_N}{\varepsilon_0}\right)\;.
\ee

We consider now the equations of motion arising from the interaction term confined to the superfluid helium volume,
\begin{equation}\label{fluidaction}
    S_{int}~=~\int\D t\D^2\bm{x}\int_0^{h(t,\bm{x})}\D z\,\frac{\alpha\rho_N}{2}\left(\frac{A_0\omega_L}{\sqrt{2}}+\partial_t\psi\right)^2\,.
\end{equation} Only the linear contribution to \eqref{fluidaction} is relevant for the interaction as the quadratic part contributes to the effective speed of light \eqref{effectiveC}. Varying the linear term with respect to the phase field and momentarily dropping the argument of $h(t,\bm{x})$ yields~\cite{luke_1967},
\begin{align}
    \delta S_{int}~=~&\int\D t\D^2\bm{x}\int_0^h\D z\,\frac{\alpha\rho_N}{\sqrt{2}}A_0\omega_L\p_t\delta\psi\,,\\
    =~&-\int\D t\D^2\bm{x}\,\frac{\alpha\rho_N}{\sqrt{2}}A_0\omega_L\p_t(h)\left.\delta\psi\right|_{z=h}\,,\\
    =~&-\frac{\alpha\rho_N}{\sqrt{2}}A_0\omega_L\int\D t\D^2\bm{x}\D z\,\p_t(h)\delta(z-h)\delta\psi\,,
\end{align} where, in the second inequality, we used the Leibniz integral rule with vanishing boundary terms and in the third equality, we rewrote the valuation $z=h(t,\bm{x})$ as an integral over $z$, introducing a delta function. 

We consider perturbations of the height with respect to a constant background, $h(t,\bm{x})=h_0+\delta h(t,\bm{x})$, hence to first order, we find
\begin{equation}
    \delta S_{int}~=~-\frac{\alpha\rho_N}{\sqrt{2}}A_0\omega_L\int\D t\D^2\bm{x}\D z\,(\p_t\delta h)\delta(z-h_0)\delta\psi\,.
\end{equation}
We absorb a factor of $A_0/\sqrt{2}$ into $\delta h(t,\bm{x})$ and introduce a factor of $\delta(\bm{x}-\bm{X}(t))$ to localise the interaction to the laser trajectory. We can then write the variation of the interaction part of the action as,
\begin{align}
    \delta S_{int}~=~-\alpha \rho_N\omega_L\int\D t\D^2\bm{x}\D z\,(\partial_t\delta h(t,\bm{x}))\delta\psi(t,z)
    \delta(\bm{x}-\bm{X}(t))\delta(z-h_0)\,.
\end{align} Variation of the entire action $S=S_{em}+S_{int}$ with respect to the laser phase field results in the equations of motion
\begin{align}
    -\frac{1}{\ceff^2}\partial_t^2\psi(t,z)+\partial_z^2\psi(t,z)~=~
    -\alpha\rho_N\omega_L\mu_0\partial_t\delta h(t,\bm{X}(t))\delta(z-h_0)\,.
\end{align}The homogeneous part gives the solution $\psi_0(t,z)$, whereas the inhomogeneous part can be solved using the Green's function for the operator $-\ceff^{-2}\partial_t^2+\partial_z^2$, which is $G(t,z)=\frac{\ceff}{2}\Theta(t-|z|/\ceff)$, where $\Theta(x)$ is the Heaviside function. This results in the solution
\begin{align}
\psi(t,z)~=~\psi_0(t,z)-\frac{1}{2}\alpha\rho_N\omega_L\mu_0\ceff\delta h(\tau,\bm{X}(\tau))\,,
\end{align}where $\tau=t-|z-h_0|/\ceff$.

\section{Effective field theory\label{app:eft}}
The system of equations in the isothermal limit 
\begin{align}
    \rho \partial_t \delta h +\rho h_0\nabla^2\phi  ~=~ & 0\,, \\
    \partial_t \phi + g \delta h ~=~ & 0\,, \label{hphi}
\end{align}
can be derived with the Hamiltonian
\be 
\mathcal{H} ~=~ \frac{1}{2}\rho\left[ h_0\left(\nabla\phi\right)^2+g\left(\delta h\right)^2\right]\;,
\ee 
and the equations of motion
\begin{align}
    \rho \partial_t \delta h ~=~ & \frac{\delta \mathcal{H}}{\delta \phi}\;, \\ 
    \partial_t \phi ~=~ & -\frac{1}{\rho}\frac{\delta \mathcal{H}}{\delta (\delta h)}\;.
\end{align}
Expanding the fields in modes 
\begin{align}
\delta h(t,\op{x}) ~=~ & \frac{1}{\sqrt{2V}}\sum_{k\neq 0} A_k\left(b_k\E^{-\I(\omega_k t-\op{k}\cdot\op{x})}-\mathrm{h.c.}\right)\,,\\ \label{fieldmode}
\phi(t,\op{x}) ~=~ & \frac{-\I g}{\sqrt{2V}}\sum_{k\neq 0}  \frac{A_k}{\omega_k} \left(b_k\E^{-\I(\omega_k t-\op{k}\cdot\op{x})}+\mathrm{h.c.}\right) \;,
\end{align}
with $\omega_k = c_3|\bm{k}|$ and 

\be 
A_k ~=~ \left(\frac{\hbar \omega_k }{\rho g}\right)^{1/2} \;.
\ee 
We quantise using the canonical commutation relations $\left[\hat{\phi}(t,\bm{x}),\rho\delta \hat{h}(t,\bm{x}')\right]= -\I\hbar\delta^{(2)}(\bm{x}-\bm{x}')$~\cite{PhysRevLett.46.1351}.
 Averaging $\delta \hat{h}$ over a Gaussian (laser) intensity profile  results in 
\be \label{deltah}
\hat{\overline{\delta h}}(t) ~=~ \int\!\D \op{x}^2 \delta \hat{h}(t,\op{x}) \frac{1}{2\pi r_0^2}\E^{-\frac{x^2+y^2}{2r_0^2}}\;,
\ee 
or calculating the integral explicitly 
\be 
\hat{\overline{\delta h}}(t) ~=~ \frac{1}{\sqrt{2V}}\sum_{k\neq 0} A_k\E^{-\frac{1}{2}r_0^2k^2}\left(\hat{b}_k\E^{-\I\omega_k t}-\mathrm{h.c.}\right)\,,\label{eq:avdeltah}
\ee 
The spatial averaging of $\delta h$ over the laser beam profile \eqref{deltah} effectively imposes a physical cutoff for wavenumbers higher than $k_0=2\pi/r_0$. In terms of the detector's frequency $\omega$, this cutoff corresponds to an upper bound at $\omega_0\equiv c_3 k_0 \approx (2\pi)~8.83\,\mathrm{kHz}$ for a beam radius of $10\,\mathrm{\mu m}$. In practice, we operate in a frequency regime much smaller than $\omega_0$.

The link between the hydrodynamical and quantum field theoretic (QFT) fields can be seen in by direct comparison of the QFT field expansions with the continuous limit of \eqref{fieldmode}. In the QFT framework, the quantised field $\hat{\varphi}$ in analogue spacetime with effective speed of light $c_3$ can be expanded as
\begin{equation}\label{QFTmode}
    \hat{\varphi}(t,\bm{x}) = \frac{c_3}{\sqrt{\hbar}}\int\frac{\mathrm{d}^2\bm{k}}{(2\pi)^2\sqrt{2\omega_k}}\left(\hat{a}_{\bm{k}}\E^{-\I\omega_k t+\I\bm{k}\cdot\bm{x}}+\mathrm{h.c.}\right)\,,
\end{equation} whereas the quantised velocity potential at the free surface in the continuous limit is written
\begin{equation}
    \hat{\phi}(t,\bm{x})=-\I\sqrt{\frac{\hbar g}{\rho}}\int \frac{\mathrm{d}^2\bm{k}}{(2\pi)^2\sqrt{2\omega_k}}\left(\hat{b}_{\bm{k}}\E^{-\I\omega_k t+\I\bm{k}\cdot\bm{x}}+\mathrm{h.c.}\right)\,.
\end{equation} By matching coefficients, one may then map the the fields into each other as \begin{equation}\label{matching}
    \hat{\phi}(t,\bm{x})=-\I\sqrt{\frac{\hbar^2 g}{\rho c_3^2}}\hat{\varphi}(t,\bm{x})\,.
\end{equation}

\section*{References}
\bibliographystyle{iopart-num} 
\bibliography{zzz_bibliography}

\end{document}